\documentclass[10pt,preprint2]{aastex}

\newcommand\plotonerot[1]{%
 \centering 
 \leavevmode 
 \includegraphics[angle=90,width={\columnwidth}]{#1}%
}%

\def\dim#1{\mbox{\,#1}}
\def\figdir{.}

\begin{document}

\pagestyle{myheadings}
\markright{DRAFT: \today\hfill}

\title{Linear Gas Dynamics in the Expanding Universe: Why the Jeans Scale
  does not Matter}
\author{Nickolay Y.\ Gnedin, 
  Emily J.\ Baker, Thomas J.\ Bethell, Meredith M.\ Drosback, A.\
  Gayler Harford, Amalia
  K.\ Hicks, Adam G.\ Jensen, Brian A. Keeney, Christopher M.\ Kelso, Mark
  C.\ Neyrinck, Scott E.\ Pollack, Timothy P.\ van Vliet}
\affil{Department of Astrophysical and Planetary Sciences, 
University of Colorado, Boulder, CO 80309}
\email{gnedin@casa.colorado.edu, ejbaker@colorado.edu, thomas.bethell@colorado.edu,
  meredith.drosback@colorado.edu, gayler\_harford@post.harvard.edu,
  amalia.hicks@colorado.edu, adam.jensen@colorado.edu,
  brian.keeney@colorado.edu,
  kelsoc@colorado.edu, neyrinck@glow.colorado.edu, pollacks@colorado.edu,
  timothy.vanvliet@colorado.edu}

\begin{abstract}
 We investigate the relationship between the dark matter and baryons in the
 linear regime. This relation is quantified by the so-called ``filtering
 scale''. We show that a simple gaussian ansatz which uses the filtering
 scale provides a good approximation to the exact solution.
\end{abstract}


\section{Introduction}

If humans had dark matter vision, cosmology would be a solved problem by
now. However, this is not the case, and in our attempt to understand the
distribution of matter in the universe we need to rely on baryons - stars
and gas - that trace the underlying distribution of the dark matter.

The baryons, though, are not the perfect tracer, since they are subject to
other forces in addition to gravity. In this paper we restrict our
attention to the
linear regime, but even in this simplest physical situation the
relationship between the dark matter and baryons (i.e.\ cosmic gas) is
nontrivial, because gas pressure will erase fluctuations on small
scales.

The effect of gas pressure on small fluctuations is canonically
described by the Jeans scale, which is defined as the scale at which
the gravity force equals the gas pressure force. On large scales 
gravity wins, and small fluctuations grow exponentially, while on small
scales, gas pressure turns all fluctuations into sound waves.

However, in the expanding universe the Jeans scale becomes essentially
irrelevant, because gravitational instability leads only to slow, power-law
growth of fluctuations. Let us consider the following simple thought
experiment: in a universe with linear dark matter fluctuations,
the gas is instantaneously heated to high temperature. The Jeans scale also
increases by a large factor instantaneously, whereas it takes about a
Hubble time for the fluctuations in the gas to respond to the changed
pressure force. Thus, the instantaneous value of the Jeans scale does not
correspond to the characteristic scale over which the fluctuations are
suppressed, but instead the suppression scale - which we call the
``filtering'' scale following Gnedin \& Hui (1998) - depends on the whole
previous thermal history. Only in the unphysical case of temperature
evolving as an exact power-law of the scale factor at all times does the
Jeans scale become proportional to the filtering scale (Bi,
Borner, \& Chu 1992; Fang et al.\ 1993). However, incorrect expressions for
the pressure force filtering have been used even until quite recently
(c.f.\ Choudhury, Padmanabhan, \& Srianand 2001)

In this paper we investigate the role of the filtering scale
further. Specifically, it has been suggested (Gnedin \& Hui 1998; Gnedin
1998) that the relationship between the gas density fluctuation $\delta_B$
and the dark matter fluctuation $\delta_X$ in the Fourier domain can be
approximated by the following simple expression:
\begin{equation}
  {\delta_B\over\delta_X} = e^{\displaystyle-k^2/k_F^2},
  \label{app}
\end{equation}
where $k$ is a wavenumber, and $k_F$ is the wavenumber corresponding to
the filtering scale. Our goal will be to investigate the range of validity
of this approximation.

\section{Linear Dynamics in the Expanding Universe}

The linear evolution of fluctuations in the expanding universe containing 
dark matter and cosmic gas is described by two second-order differential
equations (Peebles 1980),
\begin{eqnarray}
        {d^2\delta_X\over dt^2} + 2H{d\delta_X\over dt} & = &
        4\pi G\bar\rho(f_X\delta_X+f_B\delta_B),\nonumber\\
        {d^2\delta_B\over dt^2} + 2H{d\delta_B\over dt} & = &
        4\pi G\bar\rho(f_X\delta_X+f_B\delta_B) \nonumber\\
	& & \phantom{AAAA}- 
        {c_S^2\over a^2} k^2 \delta_B, 
        \label{let}
\end{eqnarray}
where $\delta_X(t,k)$ and $\delta_B(t,k)$ are Fourier components of 
density fluctuations 
in the dark matter 
and cosmic gas,
which have respective mass fractions $f_X$ and 
$f_B$, $H(t)$ is the Hubble constant, $a(t)$ is the cosmological
scale factor, $\bar\rho(t)$ is the
average mass density of the universe, $c_S(t)$ is the sound speed in the
cosmic 
gas (where the sound speed is simply defined by $c_S^2 \equiv dP/d\rho$, 
assuming an equation of state that relates the $P$ and $\rho$), $k$ is the
comoving wavenumber and $t$ is the proper time.  

On large scales ($k\rightarrow0$) the relationship between the fluctuations
in the gas and in the dark matter can be expanded as
\begin{equation}
        {\delta_B(t,k)\over\delta_X(t,k)} = 1-{k^2\over k_F^2} +
        O(k^4),
        \label{defkf}
\end{equation}
where $k_F$ is the filtering scale and in general is a function of time. 

The filtering scale is related to the Jeans scale $k_J$,
\begin{equation}
        k_J \equiv  {a \over c_S}\sqrt{4\pi G\bar\rho},
        \label{defkj}
\end{equation}
by the following relation:
\begin{eqnarray}
        {1\over k_F^2(t)} & = & {1\over D_+(t)} \int_0^t dt^\prime 
        {\ddot{D}_+(t^\prime)+2H(t^\prime)\dot{D}_+(t^\prime) 
        \over k_J^2(t^\prime)} \nonumber\\
        & & \phantom{AAAAA}
        a^2(t^\prime)
	\int_{t^\prime}^t{dt^{\prime\prime}\over a^2(t^{\prime\prime})},
        \label{kfaskj}
\end{eqnarray}
where $D_+(t)$ is the linear growing mode in a given cosmology (Gnedin \&
Hui 1998).

Inspection of equation (\ref{kfaskj}) shows that the filtering scale {\it
as a function of time\/} is related to the Jeans scale {\it as a function of 
time\/}, but at {\it a given moment in time\/} those two scales are
unrelated and can be 
very different. Thus, given the Jeans scale at a specific moment in time,
nothing can be said about the scale over which the fluctuations in the gas
are suppressed. It is only when the whole time evolution of the Jeans
scale up to some moment in time is known that 
the filtering scale at this moment can be uniquely defined.
  
\begin{figure}
\plotone{\figdir/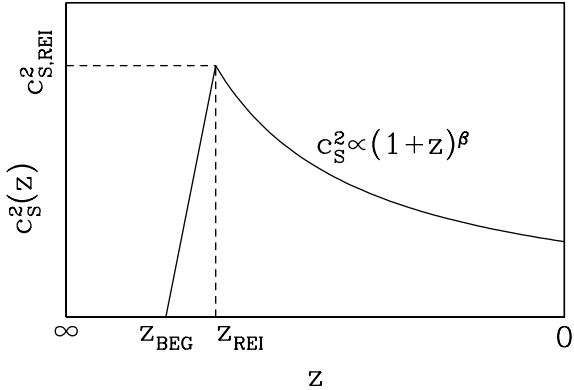}
\caption{\label{figPS}Schematic representation of our parametrization of
  the thermal history.}
\end{figure}
In order to investigate the properties of equations (\ref{let}) as a
function of the thermal history of the universe, we parameterize the
evolution of the sound speed (as a function of cosmological redshift $z$)
in the following way. Before the beginning of reionization at 
$z=z_{\rm BEG}$ the sound speed is much smaller than the sound speed in the
photoionized intergalactic medium (IGM), and is set to zero. Between 
$z_{\rm BEG}$ and $z_{\rm REI}$ the temperature in the IGM rises linearly
with redshift. The sound speed reaches a maximum at the moment of
reionization $z_{\rm REI}$ with the value of $c_{S,{\rm REI}}$, after which
moment it falls off as $(1+z)^{\beta/2}$. This parametrization is
schematically illustrated in Figure \ref{figPS}. Thus, our thermal history
is parametrized with 4 parameters: $z_{\rm BEG}$, $z_{\rm REI}$, $c_{S,{\rm
    REI}}$, which we replace with the mean temperature at reionization in
units of $10^4\dim{K}$, $T_{4,\rm REI}$, and $\beta$.
 We adopt a flat cosmological model as our
background cosmology, which introduces two more parameters: $\Omega_0$ and
$\Omega_B$. We adopt the following values as our fiducial model:
\begin{eqnarray}
  \Omega_0 & = & 0.3 \nonumber\\
  \Omega_B & = & 0 \nonumber\\
  T_{4,\rm REI} & = & 2.5 \nonumber\\
  z_{\rm REI} & = & 7 \nonumber\\
  z_{\rm BEG} & = & 10 \nonumber\\
  \beta & = & 1.
\end{eqnarray}

\begin{figure}
\plotone{\figdir/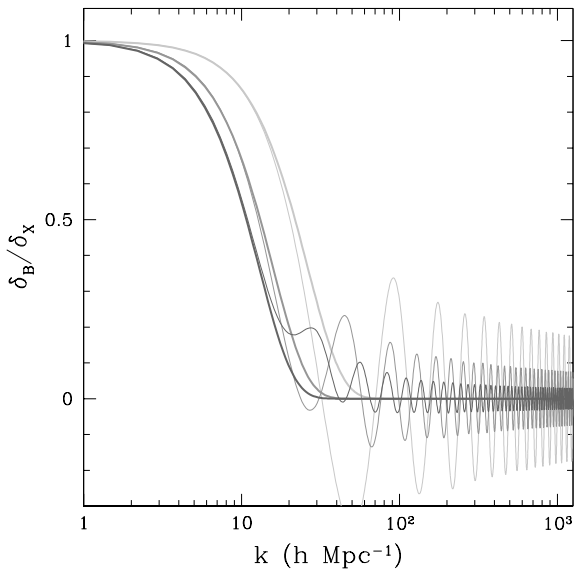}
\caption{\label{figLS}Solutions to equations (\protect{\ref{let}}) for our
fiducial model at $z=4$ (light grey), $z=1.5$ (medium grey), and $z=0$
(dark grey): thin lines show the exact solutions, thick lines give the
approximation (\protect{\ref{app}}).}
\end{figure}
Solutions to equations (\ref{let}) at three different redshifts are shown
in Figure \ref{figLS} with thin lines. As can be expected, on large scales
($k\rightarrow0$) the gas follows the dark matter, whereas on small scales
oscillations in the gas turn into slowly decaying sound waves. We also show
with thick lines the gaussian 
approximation (\ref{app}) to illustrate its level
of accuracy for our fiducial model. Clearly, the gaussian approximation
does not reproduce the small scale oscillations, but it appears to do a
decent job in reproducing the transitional region between large and
small scales.

\section{Results}

In this section we present our investigation of the accuracy of the gaussian
approximation (\ref{app}). Our approach to testing the approximation is
motivated by its applicability. We can envision using the approximation as
a simpler way of computing the rms density fluctuations in the gas:
$$
  \sigma^2_B(t) = {1\over 2\pi^2} \int_0^\infty P_X(k)
  \left[\delta_B(t,k)\over\delta_X(t,k)\right]^2 k^2dk  \approx
$$
\begin{equation}
  \approx {1\over 2\pi^2} \int_0^\infty P_X(k)
   e^{\displaystyle-2k^2/k_F^2(t)}k^2dk,
   \label{sig}
\end{equation}
where $P_X(k)$ is the dark matter power spectrum. Another use of the
approximation could be in giving the shape of the gas power spectrum on small
scales: 
$$
  P_B(k,t) = P_X(k) \left[\delta_B(t,k)\over\delta_X(t,k)\right]^2  \approx
$$
\begin{equation}
  \approx P_X(k)
   e^{\displaystyle-2k^2/k_F^2(k)}.
   \label{pow}
\end{equation}
Thus, our goal is to estimate the accuracy of the gaussian approximation
both in equation (\ref{sig}) and in equation (\ref{pow}) in our
six-dimensional parameter space. 

A full sampling of this parameter space would be unrealistic, so we focus on
varying two of the parameters at a time, keeping the rest of the parameters
fixed to our fiducial values. For the dark matter power
spectrum $P_X(k)$, we adopt a canonical Cold Dark Matter + Cosmological
Constant power spectrum with $\Omega_0h=0.25$, Harrison-Zel'dovich
scale-free primordial spectrum, and the BBKS transfer function (Bardeen et
al.\ 1986). 

\begin{figure}
\plotonerot{\figdir/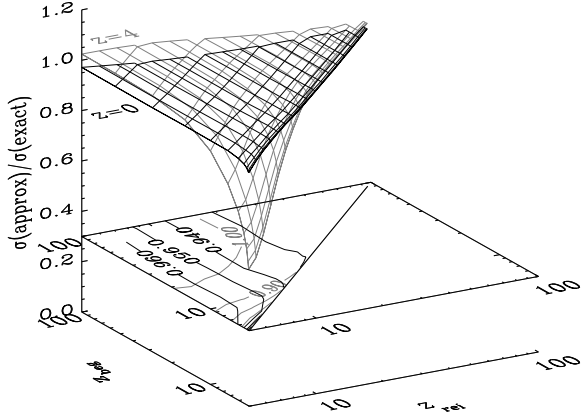}
\caption{\label{figBN}The accuracy of the gaussian approximation in
computing the rms gas fluctuation as a function of $z_{\rm REI}$ and
$z_{\rm BEG}$ at $z=4$ (grey) and $z=0$ (black).}
\end{figure}

\begin{figure}
\plotonerot{\figdir/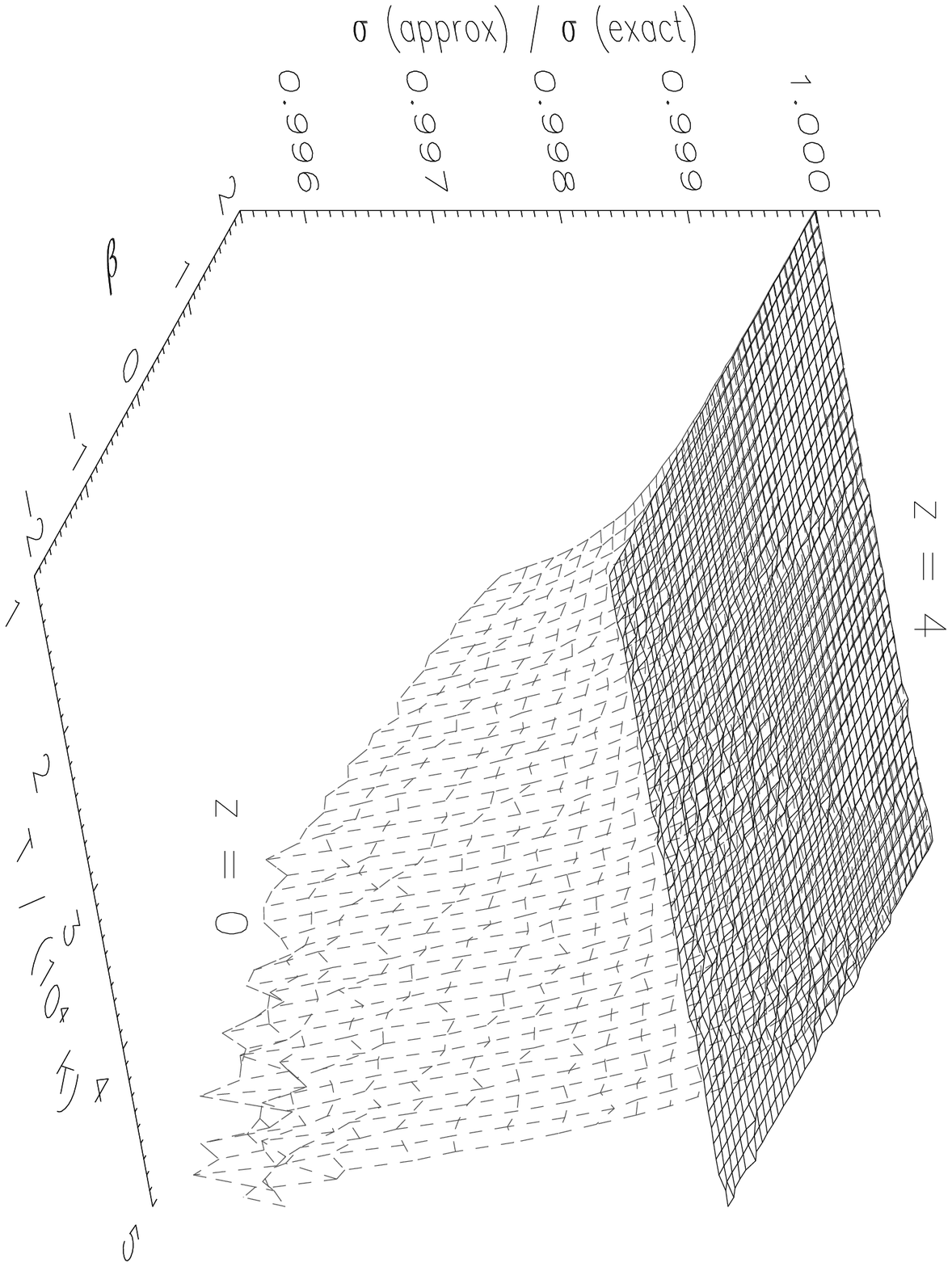}
\caption{\label{figPV}The same as Fig.\ \protect{\ref{figBN}}, except
as a function of $T_{4,\rm REI}$ and $\beta$.}
\end{figure}

\begin{figure}
\plotonerot{\figdir/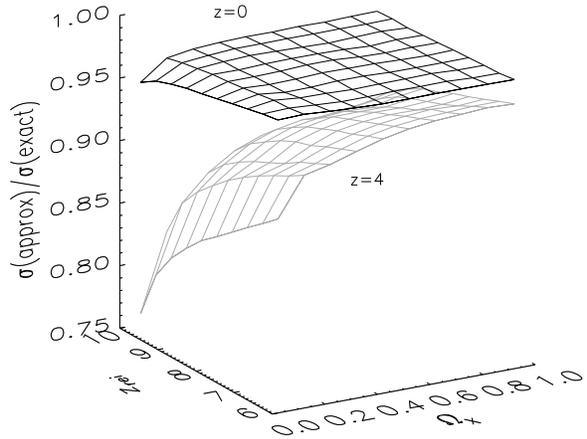}
\caption{\label{figBK}The same as Fig.\ \protect{\ref{figBN}}, except
as a function of $\Omega_X$ and $z_{\rm REI}$ (with $z_{\rm BEG}$ begin
kept fixed at 10).}
\end{figure}

\begin{figure}
\plotonerot{\figdir/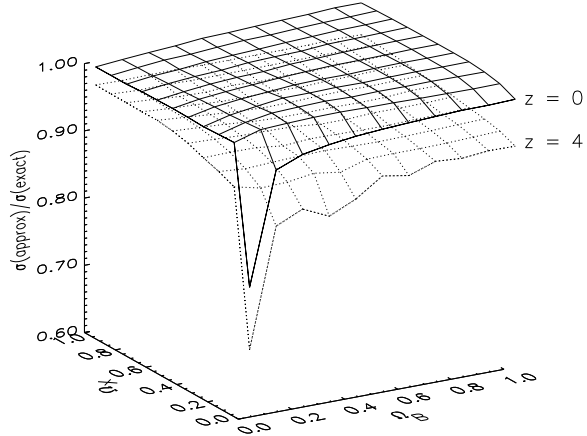}
\caption{\label{figHK}The same as Fig.\ \protect{\ref{figBN}}, except
as a function of $\Omega_X$ and $\Omega_B$.}
\end{figure}

\begin{figure}[ht]
\plotonerot{\figdir/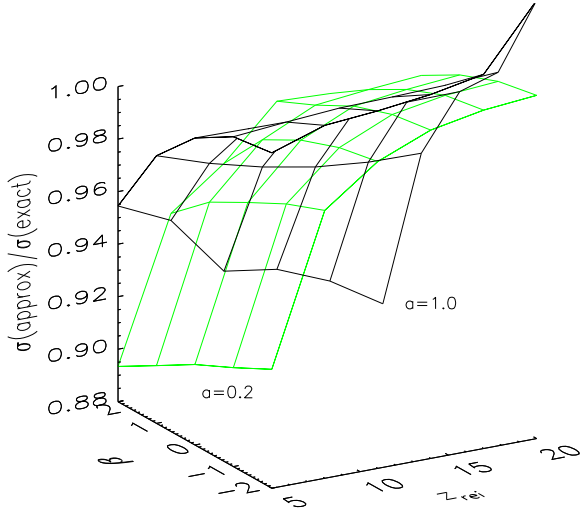}
\caption{\label{figDH}The same as Fig.\ \protect{\ref{figBN}}, except
as a function of $z_{\rm REI}$ and $\beta$.}
\end{figure}

The choice of the power spectrum does matter. For example, if the power
spectrum contains a lot of power on very small scales (below the filtering
scale), then oscillations shown in Fig.\ \ref{figLS} will be amplified,
which, in turn, will compromise the approximation (\ref{app}). However, we
believe that the dark matter power spectrum is known sufficiently well up
to scales of interest, so our choice is justified. 

Figures \ref{figBN}-\ref{figHK} show the accuracy of the gaussian
approximation as a function of our parameters. One can see that in general
the approximation holds very well, with the approximate $\sigma_B$ being
within 10\% of the exact value. The approximation breaks down at $z\ga4$
in the limit
when reionization is quite fast ($z_{\rm BEG} \la 1.2 z_{\rm REI}$) {\it
  and\/}  the redshift of reionization is low ($z_{\rm REI}<6$). It also
gets somewhat worse for very low values of $\Omega_0$. This is not
surprising, since in those two regimes the oscillations at large $k$ are
substantial and make a non-negligible contribution to the integral
(\ref{sig}). 

We have also found that in essentially all cases the gaussian approximation
gives a better than 20\% (10\% in amplitude) fit to the gas power spectrum
(\ref{pow}) for $k<0.9k_F$.

\section{Conclusions}

We have shown that the gaussian approximation (\ref{app}) provides a
reasonably good fit (better than 10\%)
to the rms density fluctuations in the gas, and for the
shape of the gas power spectrum for $k<0.9k_F$. It is important to underscore
that there is no known 
physical reason why this approximation works so well, so
it should be considered as a mathematical coincidence. Notwithstanding, the
gaussian approximation can be used in semianalytical models of the
Lyman-alpha forest and early universe, when a full solution to equations
(\ref{let}) is not practical.

\end{document}